\documentstyle[prd,aps]{revtex}

\begin{document}
\draft

\title{Exotic spacetimes, superconducting strings with linear momentum, and (not quite) all that} 
\author {Reinaldo J. Gleiser\thanks{Electronic address: gleiser@fis.uncor.edu} \and Manuel H. Tiglio\thanks{Electronic address: tiglio@fis.uncor.edu}}
\address{Facultad de Matem\'{a}tica, Astronom\'{\i}a y F\'{\i}sica,
 Universidad Nacional de C\'{o}rdoba. Ciudad Universitaria. 5000 C\'{o}rdoba, Argentina.}
\maketitle

\begin{abstract}
We derive the general exact vacuum metrics associated with a stationary
(non static), non rotating, cylindrically symmetric source. An analysis of
the geometry described by these vacuum metrics shows that they contain
a subfamily of metrics that, although admitting a consistent time
orientation,   display  ``exotic'' properties, such as ``trapping'' of
geodesics and closed causal curves through every point. The possibility that  such spacetimes could be generated by a superconducting string, endowed with a neutral
current and momentum, has recently been considered by Thatcher and Morgan. Our results, however, differ from those found by Thatcher and Morgan, and the discrepancy is explained. We also analyze the general possibility of constructing physical sources for the exotic metrics, and find that, under certain restrictions, they must always violate the dominant energy condition (DEC). We illustrate our results by explicitly analyzing the case of concentric shells, where we find that in all cases the external vacuum metric is non exotic if the matter in the shells satisfies the DEC. 
\end{abstract}


\section{Introduction}

The possibility that the spacetime around a superconducting cosmic
string with constant momentum is endowed with exotic properties has
been indicated in a recent article by Thatcher and Morgan \cite{tha}.
In accordance with the analysis carried out in \cite{tha}, in the
resulting string spacetime, test particles are deflected as they
approach the string, effectively isolating the defect from the outside
universe.  This, and other strange properties of the metric, such as
the possibility of causal violations, would imply that the inclusion of
gravity into models of charged strings and vortons may have significant
consequences as regards their possible role in a cosmological context.

Because of the very peculiar nature of the results obtained in
\cite{tha}, and since the analysis is carried out, at least in
part, through approximations or numerical integration of some complex
system of equations, it seems appropriate to try, as a check, to
reobtain those results through a different approach, and also to extend
the analysis of the properties of the associated metrics, as regards
their causal and other properties. Therefore, in this paper we start,
in Section II, with a derivation of the general form of the vacuum
metrics external to an infinite, non rotating,
axisymmetric stationary cylinder. Depending on the choice of the integration
constants that appear in solving Einstein's equations, these turn out
to be members of either two families of exact solutions of the form
of the Lewis metrics, one of which can be interpreted as a ``boosted''
Levi-Civita metric (static metrics), and the other one is similar (but
not quite equal) to the ``exotic'' metrics found in \cite{tha}
(stationary but not static metrics).

The ``boosted'' Levi-Civita metrics can be brought to a standard
diagonal form by a coordinate transformation, so the spacetime they
describe has ordinary properties. In the case of the ``exotic''
metrics this is not possible, and we find that non-spacelike {\em
geodesics} are always trapped, i.e., their radial coordinate cannot
take arbitrarily large (or small) values. Moreover, we show that although it is
possible to assign a well defined time orientation, so that we may
distinguish between future and past directed causal curves at any
point, there are causal curves through every point in these spacetimes,
whose extensions to the past and to the future eventually
self-intersect,  and that it is possible to construct non-spacelike
{\em curves} joining {\em any} two points in that spacetime. In
particular, there are closed timelike curves connecting {\em any} pair
of points in the spacetime. Since the metrics are singular on the
symmetry axis, we might expect that the singularity, as in the case of
some of the Levi-Civita metrics, can be replaced by a cylindrical
source, satisfying regularity and other physical requirements. We would
therefore have a physically admissible source for an ``exotic'' spacetime.
That such a non causal behavior results from a physical source cannot
be dismissed in principle, since it is well known that closed timelike
curves appear in spacetimes containing cosmic strings in relative
motion\cite{got}.  Unfortunately, as we show in Section III, the
construction given in \cite{tha} does not give a positive answer to the
question of the existence of such physical sources for the exotic
metric. As we indicate in that Section, the ansatz given in \cite{tha}
can be solved, once a simple coordinate transformation is introduced,
by a diagonal metric, and therefore there is no exotic behavior.
Basically, since the momentum is constant, one can always choose
a new coordinate frame in which it is zero.

In Section IV we extend the results of \cite{gle}, relating  properties
of the metric coefficients to those of the source, to the non diagonal
case, the main result being that a source for the ``exotic'' metrics
must violate the dominant energy condition (DEC) under quite general
conditions. In Section V we exemplify these results by considering the
possibility of having an ``exotic'' metric outside one or more
concentric cylindrically symmetric shells with regular axis, and
showing explicitly that this requires that the DEC is violated by the matter making up the shells.

We use geometrized units ($G=c=1$), the signature of the metric is $(-++++)$, and (sometimes) we use abstract indexes.

\section{Cylindrically symmetric, non rotating, stationary vacuum metrics}
 
We consider first the vacuum spacetime external to a non rotating, cylindrically symmetric distribution of mass-energy, in a stationary state of motion along the symmetry axis. This may be described, in general, by a metric of the form 
\begin{equation}
ds^2=  dr^2 + g_{\theta \theta} d\theta^2 + g_{zz} dz^2 + 2 g_{tz} dz\; dt - g_{tt} dt^2 ,
\label{met1}
\end{equation}
where $g_{\theta \theta} ,g_{zz},g_{tt},g_{tz}$ are functions of $r$. The general solution of the vacuum Einstein equations corresponding to (\ref{met1}) may be written in the form
\begin{eqnarray}
\label{coef1}
g_{\theta \theta} & = & C_5 (r-r_0)^{2 k1} \nonumber \; ,  \\
g_{zz} & = &  (C_1)^2 |r - r_0|^{2 k2}- (C_3)^2 |r - r_0|^{2 k3} \label{coef2} \;,  \\
g_{tt} & = &   -(C_2)^2 |r - r_0|^{2 k2} + (C_4)^2 |r - r_0|^{2 k3} \;, \label{coef3} \\
g_{tz} & = &   C_1 C_2|r - r_0|r^{2 k2}- C_3 C_4 |r - r_0|^{2 k3} \;,  \label{coef4}
\end{eqnarray}
where $r_0$ is a real constant, and the constants $k_i$ ($i=1\ldots 3$) must satisfy
\begin{equation}
\label{condk}
k_1+k_2+k_3=1 \;\;, \;\; (k_1)^2+(k_2)^2+(k_3)^2=1 \; .
\end{equation}
We must also choose the constants $C_i$, such that
\begin{equation}
\label{condC}
C_1 C_4 \neq C_2 C_3  \; ,
\end{equation}
otherwise the metric would be degenerate. We shall assume that $\theta$
is restricted to $0 \leq \theta \leq 2 \pi$, with $\theta =0$ and
$\theta =2 \pi$ identified, consistent with the interpretation of
cylindrical symmetry. In this context $r$ and $z$ would, in
principle, be the remaining ``cylindrical coordinates'', but,  as we
show below, this needs closer examination in general.

For any choice of {\em real} $k_i$, $C_i$ satisfying equations (\ref{condk})
and (\ref{condC}), we may parameterize the constants $k_i$ by a real
parameter $\Delta$,
$$
k_1 =  - \frac{2(\Delta -1)}{(\Delta ^2 +3)} \;\;\;\; , \;\;\;\; k_2 =\frac{2(\Delta +1)}{(\Delta ^2+3)} \;\;\; , \;\;\;\; k_3 = \frac{(\Delta ^2-1)}{(\Delta ^2+3)} \; ,
$$
and introduce a linear transformation of the coordinates $(z,t)$ of the form
\begin{equation}
\label{transf1}
z \longrightarrow C_1 z +C_2 t \;\;, \;\; t \longrightarrow C_3 z +C_4 t ,
\end{equation}
that puts the metric (\ref{met1}) in the usual diagonal Levi-Civita form,
$$
ds^2=  dr^2 + p_1 (ar - r_0)^{-4(\Delta -1)/(\Delta ^2 +3)} d\theta^2 + p_2 (ar - r_0)^{4(\Delta +1)/(\Delta ^2+3)}  dz^2 - p_3 (ar - r_0)^{2(\Delta ^2-1)/(\Delta ^2+3)} dt^2 ,
$$
with $p_1$, $p_2$, and $p_3$ arbitrary real and positive constants. Therefore, in this case the metric given by (\ref{coef1},\ref{coef2},\ref{coef3},\ref{coef4}) is just a ``boosted'' form of the Levi-Civita metric.

We notice, however, that the Einstein equations are also satisfied if we choose $k_1$ real, and $(k_2 , k_3)$ as a complex conjugate pair, such that Eqs. (\ref{condk}) are satisfied. We must also choose
$$
C_3 = i (C_1)^* \;\;, \;\; C_4 = i (C_2)^*,
$$
where a star indicates complex conjugation, so that the resulting metric is real. After some appropriate renaming of constants, the metric can be written in the form (\ref{met1}), with the coefficients given by
\begin{eqnarray}
g_{\theta \theta} & = & c_1 |r - r_0|^{2 q_1} \label{exotic1} \; , \\
g_{zz} & = &  c_2^2 |r - r_0|^{2 q_2} \cos{\left[ 2k \ln (|r - r_0|) + 2 \phi_1) \right]} \label{exotic2} \; ,\\
g_{tt} & = &  - c_3^2 |r - r_0|^{2 q_2} \cos{\left[ 2k\ln (| r - r_0|) + 2 \phi_2 \right]} \label{exotic3} \; ,\\
g_{tz} & = & c_2 c_3 |r - r_0|^{2 q_2}\cos{\left[ 2k \ln (|r - r_0|) + \phi_1 +\phi_2 \right] }  \label{exotic4} \;,
\end{eqnarray}
where $\alpha$ , $k$, $c_1$, $c_2$, $c_3$, $\phi_1$, and $\phi_2$ are
arbitrary real constants, and
$$
q_1= \frac{1}{3}\left[ 1+2s(1+3k^2)^{1/2} \right]  \;\;,\;\; q_2 = \frac{1}{3}\left[ 1- s(1+3k^2)^{1/2} \right] \;\;\;\; , \;\;\;\; s = \pm 1 \; .
$$
The determinant of the metric is
$$
 c_1 c_2 c_3 (r-r_0)^2 \left[ \cos{(2\phi _1 - 2\phi _2)} -1 \right]  \; , 
$$
and, thus, the metric is nondegenerate if and only if (for $r\neq r_0$)
\begin{equation}
c_1,c_2,c_3 \neq 0 \;\;\;\;\;\mbox{and}  \;\;\;\; \phi _ 1 - \phi _2 \neq n \pi \;\;\;\; (n \in {Z})   \label{deg} \; .
\end{equation}
It is static only when $k=0$, corresponding to a Levi-Civita spacetime with $\Delta=\pm1$ or $\pm 3$ if $s=1$ or $s-1$, respectively (Levi-Civita metrics with opposite values of $\Delta$ are isometric). 

This type of metric (satisfying Eq. (\ref{deg}), and with $k \neq 0$), which, following \cite{tha}, we shall call
``exotic'' hereafter,  is of Lewis type \cite{lew}, and has
been usually analyzed in relation with rotating cylinders (i.e., the
non vanishing crossed coefficient of the metric is $g_{t\theta }$)
\cite{fre,cle}. The procedure here used to obtain these solutions is
similar to that used to obtain the ``windmill solutions'' of Macintosh
\cite{mac}.  

\subsection{Some properties of the exotic metrics}

To justify the name ``exotic metric'' we consider here (see also
\cite{tha}) several classes of non spacelike curves, both geodesics and non
geodesics. We first notice that by appropriate rescalings and linear
transformations of coordinates we may write the metric in the form
\begin{equation}
\label{formsimp}
ds^2 = dr^2 + r^{2 q_1} d\theta ^2 + r^{2 q_2} \left[
\cos{(2k\ln{r})} dz^2 - \cos{(2k\ln{r})}dt^2 + 2 \sin{(2k\ln{r})} dtdz
\right] \; ,
\end{equation}
and, therefore, the constants $c_i$ and $\phi_i$ are
irrelevant, as far as the geometrical properties of the metric are
concerned. Note, from Eq. (\ref{formsimp}), that the sign of $k$ can be freely chosen; we will take advantage of
this freedom later in Sections IV and V.
 
As already noticed in \cite{tha}, for a metric of the form
(\ref{formsimp}), the Killing vectors $\partial_t$ and $\partial_z$
change from spacelike to timelike and vice versa as one moves along the
$r$ coordinate (for fixed $\theta$). In other words, if, for fixed
$\theta$, we consider a fixed coordinate grid, where $(r,t,z)$
are taken  as cartesian coordinates, the local light cones appear to
``rotate'' along the $r$-axis, while they have fixed directions in a
given $(t,z)$ ``plane'', (i.e. a constant $(r,\theta )$ surface).
Nevertheless, the form (\ref{formsimp}) for the metric immediately
shows that, since the light cones are well defined everywhere (except,
of course, for $r=0$), we may, in spite of this ``rotation'' of the
light cones, impose a definite time orientation on the spacetime, by
simply defining a future direction at a given point, and then extending
this definition by continuity to all other points.

A definite time orientation on the spacetime, however, does not preclude the possibility of the existence of closed causal (i.e. non spacelike) curves. Some peculiar behavior regarding null curves was already noticed in \cite{tha}. We therefore start our analysis by considering geodesic curves. Cylindrical symmetry implies that there is a subfamily of geodesics with constant $\theta$. Restricting to this type of geodesics, we may write their tangent vector in the form
$$
 u^a = u^z (\partial _z)^a + u^t ( \partial _t )^a
 + u^r ( \partial _r)^a \; .
 $$

Since $ (\partial_t )^a$ and $(\partial_z)^a$ are Killing vectors, and $u^a$ is geodesic, we have
$$
 u_a (\partial _z)^a  = p \;\;,  \;\; u_a (\partial _t)^a = - E  \; ,
$$
where $E$ and $p$ are constants. From these relations, we find 
\begin{eqnarray}
u^z & = & \dot{z} =  (E^2+p^2)^{1/2}r^{- 2q_2} \cos{(2k\ln{r} - a)} \; ,  \label{geo1}  \\
u^t & = & \dot{t} =  (E^2+p^2)^{1/2}r^{- 2q_2} \sin{(2k\ln{r} - a)} \; , \label{geo2} \\
(u^r)^2 & = & \dot{r}^2= - U(r) \;\;\;\; , \;\;\; U(r) = s - (E^2+p^2) r^{- 2q_2} \cos{(2k\ln{r} -2a +\pi)} \; ,  \label{geo3}
\end{eqnarray}
where $a = \arctan(E/p)$, $s = 0$ ($s=1$) for null (timelike) geodesics, and a dot indicates derivation with respect to the affine parameter, e.g. $\dot{z}=dz/d\tau  $. 

For null geodesics, we may rescale $\tau$ to set $(E^2+p^2)^{1/2}=1$, and therefore, without loss of generality we may then write
\begin{equation}
\label{dr}
\frac{dr}{d \tau } = \pm r^{-q_2/2} \left[\cos {(2k\ln{r} -2a +\pi)} \right]^{1/2}  \; ,
\end{equation}
where the plus/minus sign corresponds, respectively, to ``outgoing'' and ``ingoing'' geodesics. Notice that this implies that {\em all} null geodesics have ``turning points''
in $r$ (points where $\cos{(2k\ln{r}-2a+\pi)}=0$), and therefore,
the corresponding values of $r$ are restricted to a finite segment of
the $r$-axis, where $\cos {(2k\ln{r} -2a +\pi)} \geq 0$. Upon reaching
a turning point the sign of $ dr /d \tau $ is changed, and the
sense of the motion along the $r$-axis is reversed, but there is no
associated singularity in the metric. Turning points are properties of individual
null geodesics, i.e., they change when we consider different null
geodesics going through the same point, and therefore there are no
horizons, or any other peculiar geometrical property associated to the
turning points of a given null geodesic.

Since null geodesics going through a certain $r$ eventually come back to the same value of $r$, it is important to compute the corresponding change in $t$ and $z$. These may be obtained by eliminating the affine parameter $\tau$, and looking essentially at the corresponding trajectories in spacetime. From equations (\ref{dr}) and (\ref{geo1},\ref{geo2}\ref{geo3}), we have
\begin{eqnarray}
\label{geo04}
{d z \over d r} & = & \pm r^{-q_2/2} {\cos{(2k\ln{r} - a)} \over \sqrt{\cos{(2k\ln{r} -2a +\pi)} }}  \; ,  \\ 
{d t \over d r}  & = & \pm r^{-q_2/2} {\sin{(2k\ln{r} - a)} \over \sqrt{\cos{(2k\ln{r} -2a +\pi)} }} \label{geo04b}\; ,
\end{eqnarray}
where the plus/minus sign is determined from Eq. (\ref{dr}). After an integration by parts of the right hand sides of Eqs. (\ref{geo04}) and (\ref{geo04b}), we obtain the following expressions for the changes in $z$ and $t$, corresponding to a null geodesic that goes from an initial point in $r=r_0$ to a turning point at $r=r_1$, and comes back to $r=r_0$,
\begin{eqnarray*}
\Delta t & = &  2[ t(r_1) - t(r_0) ] = \pm \left[ 2\alpha {\cal I} - 2k^{-1}r_0h(r_0) \cos{a} \right] \; , \\
\Delta z &= & 2[ z(r_1) - z(r_0) ] = \pm \left[ 2\beta {\cal I} + 2k^{-1}r_0h(r_0) \sin{a} \right] \; ,
\end{eqnarray*}
with
$$
{\cal I} =  \int_{r_0}^{r_1}  h(r) dr  \;\;\;\;\; , \;\;\;\; h(r) = r^{-q_2/2}\cos^{1/2}{(2k\ln{r} - 2a + \pi)} \; ,
$$
and
$$
 \alpha = - \sin{a}  - \mu \cos{a} \;\;\; , \;\;\; \beta  =- \cos{a} + \mu \sin{a} \;\;\;\; , \;\;\;\; \mu =  \frac{1}{2k}\left(-q_2+2 \right) \; .
 $$

To see if the end point can be in the causal past of the initial point, using the same arguments as those leading to the form (\ref{formsimp}) for the metric, we may choose, without loss of generality, $r_0=1$, and assume that a future directed vector at $r=1$ has $\dot{t} > 0$. Then, the end point will be in either the causal past or future of the starting point if $\Delta l ^2 = -\Delta t^2 + \Delta z ^2 \leq 0$. But we have 
$$
\Delta l^2 = \frac{4\dot{r}_0^2}{k^2}\left[ {\cal I} \mu k + h(r_0) \right]^2 + 4 {\cal I}^2(3\sin ^2{a} + \cos ^2{a}) - 8 |{\cal I}|\dot{t}_0 \Delta t \; ,
$$
so that a necessary condition for   $\Delta l^2 \leq 0$, is $\dot{t}_0 \Delta t > 0$,  and therefore the end point is causally connected to the starting point only if $\Delta t > 0$. We conclude that a null geodesics cannot return to its causal past, and since $\Delta t = 0$ implies $\Delta l^2 > 0$, null geodesics cannot self intersect.

We have seen that null geodesics necessarily have turning points. We may similarly show that timelike geodesics necessarily have turning points as well. Since these are not horizons, we may ask whether turning points necessarily appear also for general (i.e., non geodesic) causal curves. That this is not so can be illustrated with a simple example. Consider a curve parameterized so that its tangent vector satisfies
$$
u_r = s \;\;\;\;\; , \;\;\;\;\; u_z = r^{B\cos{b}-1 } \cos{\left[  \ln{(r^{B\sin{b}})} - b \right] }  \;\;\;\;\; , \;\;\;\;\;  u_t = r^{B\cos{b}-1 } \sin{\left[  \ln{(r^{B\sin{b}})} - b \right]} \; ,
$$
where $b$ and $B$ are related to $k$ and $q_2$ in Eq. (\ref{formsimp}) by
\begin{equation}
s = \pm 1\;\; , \;\; k = B\sin{b} \neq 0 \;\; , \;\;\ q_2 = 2 - 2B\cos{b}  \; .
\end{equation}
Without loss of generality we may choose $B>0$ and $\sin{b}>0$.
We may easily check that, for either choice of $s$, this represents a null curve, well defined for all values of $r$, where $r$ monotonically increases ($s=+1$), or decreases ($s =-1$), without turning points.

We may obtain the trajectory (worldline) corresponding to this curve  integrating $t$ and $z$ as functions of $r$. The result is 
\begin{eqnarray*}
t(r) & = & s \left\{  r^{B\cos{b} } \cos{\left[  \ln{(r^{B\sin{b}})} - b \right] } +C_1  \right\} B^{-1}  \; , \\
z(r) & = & - s \left\{  r^{B\cos{b} } \sin{\left[  \ln{(r^{B\sin{b}})} - b \right] } + C_2  \right\} B^{-1} \; , \\
\end{eqnarray*}
where $C_1$, and $C_2$ are integrations constants. Again, with sufficient generality, we may choose these constants such that $t=z=0$ for $r=1$. This corresponds to $C_1=\cos b$, $C_2=-\sin b$.  
As indicated, for fixed $s$ these curves have no turning points. We may, however, at any point $r=r_1$, match a portion of this curve with, say, $s=+1$, with a portion of a similar curve with $s=-1$, the same values of $B$ and $b$, and where we choose $C_1$ and $C_2$ in such a way that $r$, $t$, $z$, $u_t$, and $u_z$ are continuous, and $u_r$ changes sign. Since this is still a null curve, for a photon this would correspond to ``bouncing off a mirror'', whose normal points in the $r$ direction. More geometrically, we may always ``round corners'', without changing the non spacelike nature of the curve \cite{wal}, so that the sharp change in $u_r$ is used only to simplify the computations, but has no special significance.

We may compute now the changes in $t$, $\Delta t = t(r_1)-t(1)$, and $z$, $\Delta z = z(r_1)-z(1)$, in the ``round trip'' from $r=1$ to $r=r_1$, and back to $r=1$, along the resulting null curve. These are simply twice the changes in $t$ and $z$ in going from $r=1$ to $r=r_1$, i.e., $\Delta t = 2[ t(r_1) - t(1) ]$ and $\Delta z = 2[ z(r_1)-z(1) ]$. We define
$$
 y = \ln{(r^{B\sin{b}})} - b \;\;\;\;\; , \;\;\;\; x = r^{B\cos{b}} \;.
$$

We then have three possible cases, depending on the sign of $\cos(b)$:
\begin{itemize}
\item First, if $\cos{b} > 0$, the function $x(r)$ is increasing with $r$, and, therefore, the equation $z(r)=0$ has solutions for sufficiently large $r$. Moreover, for sufficiently large $r$, these zeros of $z$ correspond to $y \simeq n \pi$, since $x$ is unbounded. Then, again for sufficiently large $r$, and for the same reason, the zeros of $z$ correspond alternatively to positive or negative values of $t$. So we may always choose $r_1$ such that $z(r_1)=0$ and $t(r_1)<0$. But then, the return point on $r=1$ is in the causal past of the initial point, and we may construct a closed, everywhere future directed, non spacelike curve, by simply joining the end point and the initial point with the timelike, future directed curve whose trajectory corresponds to $r=1,z=0$.
\item In the case $\cos(b) < 0$, the function $x(r)$ increases as $r$ decreases. The same type of reasoning indicates that in this case there are points $r_1$, (with $r_1 < 1$), such that $z(r_1)=0$, with $t(r_1)<0$, and therefore we may also construct closed, nonspacelike, everywhere future directed null curves.
\item Finally, if $\cos(b)=0$, points with $z(r_1)=0$ correspond also to $t(r_1)=0$, and the return portion of the curve intersects the initial portion,  resulting directly in a closed curve.
\end{itemize}
Although, for simplicity, we used the example of a null curve, since this is not geodesic , we conclude that there are closed, non spacelike (either null or timelike), everywhere future directed curves through every point of the manifold (see, e.g., \cite{wal}). This last result follows by taking into account that the metric at any point on the manifold can be put in the form (\ref{formsimp}), in a coordinate patch where the coordinates of the point are $(r=1, t=0, z=0)$.

A further consequence, which is not difficult to prove, is that any pair of points in the manifold can be joined by a future directed, nonspacelike curve, irrespective of the order of the points (note that, by a simple extension of the previous discussion, given any point, we may reach a point arbitrarily in its past by an appropriate choice of $r_1$).

\subsection{A comment on the metric given by Thatcher and  Morgan }

For completeness, and to clarify a point related to the regularity of the metrics, let us now compare the metric (\ref{formsimp}) with the one given in \cite{tha}:
\begin{equation}
ds^2 = \cos{\left[ \frac{8m}{\delta } \ln{\left( \frac{r}{\delta} \right) }  \right]}(-dt^2 + dz^2)
+ 2 \sin{\left[ \frac{8m}{\delta } \ln{\left( \frac{r}{\delta} \right)}  \right]}dtdz
+ dr^2 +  (1-8E)r^2 d\theta ^2 \label{vac} \; .
\end{equation}
Contrary to what is stated in \cite{tha}, this is not a vacuum solution (unless $m=0$, which gives a flat metric), because the Ricci tensor has a (unique) non vanishing component:
$$
R_{rr} = 32 \frac{m^2}{\delta ^2 r^2} \; .
$$
The identification of (\ref{vac}) as a vacuum metric made in  \cite{tha} is based on the assumption that if, for a metric of the form (\ref{met1}) and a certain tensor ${\cal{T}}_{ab}$ having the same symmetries as $g_{ab}$ and whose divergence vanishes (i.e. $\nabla ^a {\cal{T}}_{ab} =0$), the equations:
\begin{equation}
R_{tt} = S_{rr} \;\;\; , \;\;\; R_{zz}  = S_{zz} \;\;\; , \;\;\; R_{tz} = S_{tz} \;\;\; , \;\;\; R_{\theta \theta}= S_{\theta \theta}  \;\;\;\; 
\left( \mbox{ with } S_{ab}=  8\pi\left( {\cal{T}}_{ab}-\frac{1}{2}g_{ab}{\cal{T}} \right)  \right)  \; , \label{part1}
\end{equation}
are satisfied, then we necessarily have $R_{rr}= S_{rr}$. In particular, this would imply that if Eqs. (\ref{part1}) are satisfied for ${\cal{T}}_{ab}=0$, then we necessarily have a vacuum solution. This, however, is incorrect. In fact, following Garfinkle  \cite{gar1}, we may  define the tensor 
$$
Q_{ab} = R_{ab} - S_{ab} \; .
$$
Then, equations (\ref{part1}) imply that
$Q_{ab}$ has nonvanishing components only in $rr$, i.e., 
$$
Q_{ab} = J \nabla _a r \nabla _b r    \; ,
$$
with $J$ a function of $r$. Using now the contracted Bianchi identities, $\nabla ^a R_{ab} = \frac{1}{2} \nabla _b R$, together with the assumption $\nabla ^a {\cal{T}}_{ab} =0$, we find 
$$
\nabla ^a Q_{ab} = \frac{1}{2} \nabla _b Q  \;\;\;\;\; (Q \equiv Q^{\;\;a} _a)      \;.
$$

For a metric of the form (\ref{met1}), this reduces to $(J\;g)'=0$, where $(-g)$ is the determinant of the metric, and, therefore, we have 
\begin{equation}
J =  k/g \; , \label{red}
\end{equation}
with $k$ a constant. Then, if the metric satisfies regularity conditions on the symmetry axis, 
\begin{equation}
g_{tt} = -1 + O(r^2) \;\;\;\; , \;\;\;\;g_{zz} = 1+ O(r^2)  \;\;\;\; , \;\;\;\; g_{\theta \theta } = r^ 2+ O(r^4) \;\;\;\; , \;\;\;\; g_{tz}=O(r^2)  \;,  \label{reg}
\end{equation}
taking the limit $r\rightarrow 0$ in Eq. (\ref{red}), we obtain $k=0$ (otherwise we would have $J \rightarrow \infty$, corresponding to a singular 
metric at the axis), i.e. $J=0$ and $Q_{ab} = 0$; but $J$ need not vanish if the metric is not regular for $r=0$.

\section{Superconducting strings with constant momentum}

As indicated above, it was suggested in \cite{tha} that exotic metrics of the type described in the previous Section may be associated with the spacetime external to a superconducting string with constant momentum. The construction given in \cite{tha} proceeds as follows. The metric is assumed to have the form   (\ref{met1}), the matter fields are written as $\Phi = Re^{i \psi}$, and $\sigma = Se^{i\phi }$, and their Lagrangean is
\begin{eqnarray}
{\cal L} _{\mbox{matter}}& = & -\frac{1}{2} \nabla ^a R \nabla_a R  -\frac{1}{2} \nabla ^a S \nabla_a S - \frac{1}{2} R^2 (\nabla _a \psi + e A_a)(\nabla ^a \psi + e A^a) - \frac{1}{2} S^2 \nabla _a \phi \nabla ^a \phi \nonumber \\ & &  - \lambda (R^2 - \eta ^2)^2 - \frac{1}{16 \pi} F_{ab} F^{ab}    -  fR^2S^2 - \frac{1}{4} \lambda _{2}S^4 + \frac{1}{2}m^2S^2  \; .  \label{lag} 
\end{eqnarray}
Then, the following ansatz for these fields is proposed in \cite{tha}: 
$$R=R(r) \;\;\; , \;\;\; \psi = \theta\;\;\; , \;\;\; A_a = \frac{1}{e} [P(r)-1] \nabla _a \theta \;\;\; , \;\;\; S=S(r) \;\;\; , \;\;\;\phi=kz-\omega t\;\; . $$  The form assumed for $\phi$ is consistent with the idea of endowing the string with a non vanishing momentum. With these assumptions the equations resulting from the Lagrangean (\ref{lag}) are
\begin{eqnarray}
\nabla _a \nabla ^a R - R \left[ 4\lambda(R^2-\eta ^2)+  (\nabla _{a} \psi + e A_a)(\nabla ^{a} \psi + e A^a) ) + 2fS^2  \right] & = & 0 \; , \\
\nabla _a \nabla ^a S - S \left[ \nabla _{a} \phi \nabla ^{a} \phi + 2fR^2 +\lambda_2 S^2 - m^2  \right] & = & 0 \; , \\
\nabla _a \left[ R^2 (\nabla ^a \psi + e A^a) \right] & = & 0  \label{campo1} \; ,  \\
\nabla _a \left[ S^2 \nabla ^a \phi \right] & = & 0  \label{campo2} \; ,  \\
\nabla ^a F_{ab} - 4\pi e R^2 (\nabla _{b} \psi + e A_b) & = & 0 \; . 
\end{eqnarray}
Now suppose, as in \cite{tha}, that $\nabla _a \phi$ is timelike at the axis, i.e., $\omega ^2- k^2 >0$. Performing the change of variables
$$
z \rightarrow ( \omega^2 - k^2)^{-1/2} (\omega z - k t)  \;\;\;\; , \;\;\;\; t \rightarrow (\omega^2 - k^2)^{-1/2} (kz - \omega t)  \; ,
$$
the conditions at the axis remain unaltered but $\phi \rightarrow (\omega^2 - k^2)^{1/2}t$. Thus, without loss of generality we may (and we do) assume $k =0$.

Eqs. (\ref{campo1}) and (\ref{campo2}) are automatically satisfied and the other three are
\begin{eqnarray}
\left( R' g^{1/2}\right)' g^{-1/2} - R\left[  \frac{1}{2} \lambda (R^2 - \eta ^2) + \frac{P^2}{g_{\theta \theta }} + 2fS^2 \right]& = & 0 \label{string1} \; , \\
\left( S' g^{1/2}\right)' g^{-1/2} - S\left[ - \frac{\omega ^2 g_{\theta \theta}g_{zz}}{g} + 2fR^2 + \lambda_2 S^2 - m^2 \right]& = & 0 \label{string2} \; , \\
\left( \frac{P' g^{1/2}}{g_{\theta \theta }}\right)' g_{\theta \theta} g^{-1/2}- 4\pi eR^2P =0 \label{string3} \; .
\end{eqnarray}
Einstein equation, $R_{ab} = 8\pi (T_{ab} - \frac{1}{2} T g_{ab})$, can be written as (for briefness, we do not write down the explicit expressions for the Ricci tensor)
\begin{eqnarray}
R_{tt} & = & -g_{tt}\left[ \frac{1}{e^2 g_{\theta \theta}}P^{' 2} - \lambda \pi (R^2 - \eta ^2)^2 - 2 \pi S^2\left(4fR^2+\lambda_2 S^2 - 2m^2 + \frac{4\omega ^2}{g_{tt}}  \right) \right] \label{rtt} \; , \\
R_{zz}& = & g_{zz} \left[ - \frac{1}{e^2 g_{\theta \theta }} P^{' 2} + \lambda \pi (R^2 - \eta ^2)^2 + 2 \pi S^2(4fR^2+\lambda_2 S^2 - 2m^2  ) \right]  \label{rzz} \; , \\
R_{rr} & = &  8\pi (R^{'2}+S^{'2}) + \lambda \pi (R^2- \eta ^2)^2 +2\pi S^2(4fR^2+\lambda _2 S^2 - 2m^2) + \frac{1}{e^2 g_{\theta \theta } } P^{' 2}  \label{rrr} \; , \\
R_{\theta \theta }& = & g_{\theta \theta } \left[ \frac{1}{g_{\theta \theta }} \left( \frac{1}{e^2}P^{' 2} + 8\pi R^2P^2  \right) + \lambda \pi (R^2 - \eta ^2)^2 + 2 \pi S^2(4fR^2+\lambda_2 S^2 - 2m^2 ) \right] \label{raa} \; , \\
R_{tz}& = & g_{tz} \left[ - \frac{1}{e^2g_{\theta \theta } } P^{' 2} + \lambda \pi (R^2 - \eta ^2)^2 + 2 \pi S^2(4fR^2+\lambda_2 S^2 - 2m^2 )  \right] \label{rtz} \;.
\end{eqnarray}

Suppose that we have a solution of Eqs.(\ref{string1}, \ref{string2}, \ref{string3} \ref{rtt},\ref{rzz},\ref{rrr},\ref{raa},\ref{rtz}) that satisfies the regularity conditions (\ref{reg}). We will now show that $g_{tz}=0$. For that purpose we first define
$$
p =  \int_0^r \left[ - \frac{h' g_{zz}g_{tt}}{2h^2} + \frac{g_{tz}}{g_{tz}'} \left( 2t_{tz} + \frac{g_{tt}'g_{zz}'}{h} \right)  \right] \;\;,\;\; \mbox{and} \;\;\;\;
q = g_{tz}'g_{\theta \theta}^{1/2}\exp{\left( \frac{-g_{zz}g_{tt}}{2h} \right) }\; ,
$$
with $h$ and $t_{tz}$ given by 
$$
h = g_{tt}g_{zz}-g_{tz}^2 < 0 \;\;\; , \;\;\;
t_{tz} = - \frac{1}{e^2 g_{\theta \theta }} P^{'2} + \lambda \pi (R^2 - \eta ^2)^2 + 2 \pi S^2(4fR^2+\lambda_2 S^2 - 2m^2 )   \; .
$$
The convergence for small $r$ of the integral that defines $p$ is guaranteed by the regularity of the metric (and its non degeneracy) and of the matter fields. With these definitions, Eq. (\ref{rtz}) implies that, for $r >0$,
\begin{equation}
qe^p = C    \label{diag} \; ,
\end{equation}
with $C$ a constant. Taking the limit $r \rightarrow 0$ in the l.h.s.  of Eq. (\ref{diag}) and using the conditions (\ref{reg}), we find that $C=0$, i.e. 
$g_{tz}=0$ $\forall r$.

Thus, we have shown that the ansatz for the fields given in \cite{tha},
plus regularity conditions on the axis $r=0$, imply that the metric can
be made everywhere diagonal, and this excludes the possibility of
``exotic'' behavior. Therefore, we must conclude that the ansatz of
\cite{tha} does not lead to a source for the ``exotic'' metrics. The
exotic behavior  observed in \cite{tha} would then have to be ascribed to
some peculiarity in their numerical procedures, possibly leading to a
failure in strictly satisfying the regularity conditions for $r=0$.

\section{Sources for the exotic metric}

In this Section we consider some general properties of cylindrically symmetric and stationary {\em regular} sources that satisfy the DEC, and which are confined to a cylinder of arbitrary radius ${\cal R}_1$. We assume that the spacetime external to the source is vacuum, and obtain restrictions on the possible form of the metrics representing that part of the spacetime.  We shall assume that in the source region, consistent with cylindrical symmetry, there exist two killing vector fields, one of them timelike and the other spacelike, which are not necessarily orthogonal to each other. We also require that these killing fields remain timelike and spacelike, respectively, i.e., that their norm does not vanish. Then, the metric can be written as
\begin{equation}
ds^2  = -e^A dt^2 + e^B dz^2 + 2D(r)e^{(A+B)/2}dtdz + dr^2 + e^C d\theta ^2 \; .  \label{met2}
\end{equation}
with $A,B,C,D$ functions of $r$. We shall see that this kind of sources cannot give rise to exotic spacetimes. The first part of the proof consists in recalling certain inequalities that the DEC imposes on the eigenvalues of $T_a^{\;\;\;b}$. We then rewrite one of these inequalities as a differential one that can be integrated and seen not to be satisfied by the exotic metrics. Without loss of generality, we assume that the latter have $k <0$. 

\subsection{DEC, eigenvalues and eigenvectors}

For metrics of the type (\ref{met1}), the non trivial equations for the eigenvalues and eigenvectors of  $T_a^{\;\;b}$ correspond to the $t-z$ sector. For the analysis of this subsection, it is convenient to choose an orthonormal basis $(\hat{t}^a , \hat{z^a})$ at the point where the analysis is carried out, i.e. $g_{tz}=0$, $g_{tt}=-1$, and $g_{zz}=1$. Then, the eigenvalues are given by
$$
\lambda _{(\epsilon)} = \frac{1}{2} \left[ \left( T_{zz} - T_{tt} \right) + \epsilon \gamma ^{1/2}\right] \;\;\;\; , \;\;\;\;  \gamma = \left( T_{zz} + T_{tt} \right)^2 - 4 T_{tz}^2
\;\;\;\; , \;\;\;\; \epsilon = \pm 1\; .
$$

The DEC states that, for any timelike and future directed $\hat{m}^a$, $n^a = -T_{ab}\hat{m}^a $ is future directed and causal. It is easy to see, writing $\hat{m}^a = \hat{t}^a \cosh{\xi} + \hat{z}^a \sinh{\xi} $, that, if the eigenvalues are complex ($\gamma < 0$), $n^a$ is always spacelike if $T_{zz} - T_{tt} =0$ , and it is spacelike for some $\xi$ if $T_{zz} - T_{tt} \neq 0$.

On the other hand, $n^a$ is future directed $\forall \xi$ if and only if
$$
n^a \hat{t}_a = (T_{zz} + T_{tt}) \sinh ^2{\xi} + T_{tt} \geq 0  \;,
$$
a condition that is satisfied $\forall \xi$ if and only if $T_{tt}\geq 0$ and $T_{zz}+T_{tt}\geq 0$. 

Suppose that the DEC is satisfied, then $\gamma \geq 0$ and there are two real eigenvalues. The norms of their corresponding eigenvectors,  $e^a_{(\epsilon )}$, are then given by
$$
e^a_{(\epsilon )}e_{(\epsilon )a} = \frac{\epsilon}{2} \gamma^{1/2} \left( T_{zz} + T_{tt} + \gamma^{1/2} \right) \; .
$$
If $\gamma = 0$, there is a unique (null) eigenvector, $\hat{t}^a+\hat{z}^a$, with multiplicity $2$, and the energy momentum tensor is of type II in the classification of \cite{haw}. If $\gamma >0$, one of the eigenvectors is timelike and the other one spacelike. In this case the stress tensor is of type I in the classification of \cite{haw}.

\subsection{The proof}
In a non orthogonal system of coordinates, such as that one of (\ref{met2}), the eigenvalues of $T_{a}^{\;\; b}$ are
$$
\lambda = (2h)^{-1} \left\{ T_{tt}g_{zz} + T_{zz}g_{tt} - 2T_{tz}g_{tz} \pm \left[ (T_{tt}g_{zz} - T_{zz}g_{tt})^2 -
4(T_{zz}g_{tz} - T_{tz}g_{zz})(T_{tz}g_{tt} - T_{tt}g_{tz})  \right]^{1/2}       \right\}  \; .
$$
The reality of the eigenvalues is thus equivalent to
\begin{equation}
\alpha \equiv (T_{tt}g_{zz} - T_{zz}g_{tt})^2 - 4(T_{zz}g_{tz} - T_{tz}g_{zz})(T_{tz}g_{tt} - T_{tt}g_{tz})  \geq 0  \label{reality}  \; .
\end{equation}
For the metric (\ref{met1}), we have
$$
\alpha =  \frac{1}{64 \pi ^2} ( \alpha _1^2 - \alpha _2^2 )  \; ,  \label{des1} 
$$
with
\begin{eqnarray}
\alpha _1 & = & \frac{1}{2}(1+D^2) e^{(A+B-C)/2} \left[ (A-B)'e^{(A+B+C)/2}(1+D^2)^{-1/2} \right]' \label{alpha1} \; , \nonumber \\
\alpha _2 & = & e^{(A+B-C)/2} (1+D^2)^{1/2}  \left[ D' e^{(A+B+C)/2}(1+D^2)^{-1/2}  \right]' - \frac{D}{4}e^{(A+B)} (A-B)'^2 \; , \nonumber
\end{eqnarray}
It can also be seen that
\begin{equation}
\alpha _1 =  8\pi (1+D^2)^{1/2}e^{(A+B)}(T_{ab}\hat{t}^a\hat{t}^b + T_{ab}\hat{z}^a\hat{z}^b)   \label{wec1} \; ,
\end{equation}
so the WEC $\Rightarrow \alpha _1 \geq 0$. Although it is not related to our proof, let us show something else. From Eqs. (\ref{alpha1}) and (\ref{wec1}) we obtain
$$
(A-B)'e^{(A+B+C)/2}(1+D^2)^{-1/2} = 16\pi \int _0^r e^{(A+B+C)/2}(1+D^2)^{-1/2}  (T_{ab}\hat{t}^a\hat{t}^b + T_{ab}\hat{z}^a\hat{z}^b) \; ,
$$
and the WEC $\Rightarrow (A-B)' \geq 0$. Recalling that $(A-B)_{r=0}=0$, we have $|g_{tt}| \geq g_{zz} \forall r$. 

Returning to our proof, the point is that the nonnegativity of $\alpha _1$ implies that (\ref{reality}) can be written as $\alpha _1 - |\alpha _2| \geq 0$. But $\alpha _1 - |\alpha _2| \geq 0 \Rightarrow \alpha _1 - \alpha _2 \geq 0$. And we have
$$
\alpha _1 - \alpha _2 = \frac{1}{2} (1+D^2)e^{(A+B-C)/2-F} \left( qe^F\right)'  \; ,
$$
with
$$
q = A' -B' - 2D'(1+D^2)^{-1/2}    \;\;\;\;\; , \;\;\;\;\; F = \int_0 ^r \frac{D \left[ (A-B)'(1+D^2)^{1/2} + 2D'\right]}{2(1+D^2)e^{(A+B)/2}}  \; ,
$$
(convergence of the integral that defines $F$ is guaranteed by Eqs. (\ref{reg})). Thus the DEC is equivalent to 
\begin{equation}
\left( qe^F\right)'  \geq 0 \; .   \label{finaldes}
\end{equation}
Integrating Eq. (\ref{finaldes}) and using once again Eqs. (\ref{reg}), we have $q \geq 0$ $\forall r$. In particular, $q \geq 0$ at the radius of matching, $r= {\cal R}_1$. {\em By regularity}, $q$, which is made up of the metric components and its first derivatives, is continuous , and, therefore, at $r={\cal R}_1$ it can be evaluated using the exotic metrics, finding
\begin{equation}
q|_{r={\cal R}_1}   =  \frac{2k \left[ (1+\sin{\alpha _1})\cos{\alpha _2} + (1-\sin{\alpha _2})\cos{\alpha _1}  \right] } {({\cal R}_1 - r_1)\cos{\alpha _1} \cos{\alpha _2} } \geq 0  \; , \label{ter2}
\end{equation}
with
$$
\alpha _1 \equiv 2k\ln{({\cal R}_1-r_1)} + 2\phi_1 \;\;\;\; , \;\;\;\; \alpha _2 \equiv 2k\ln{({\cal R}_1-r_1)} + 2\phi_2 - \pi \; .
$$
Let us now analyze the rhs of Eq. (\ref{ter2}). In order for the exterior (exotic) metric to be regular, ${\cal R}_1-r_1 >0$. We also have that $\cos{\alpha _1} >0$ and $\cos{\alpha _2} >0$, since the metric coefficients at $r=R_1$ are
$$
g_{zz} = c_2^2 ({\cal R}_1 - r_1)^{2 q_2} \cos{\alpha _1} > 0 \;\;\;\;\; , \;\;\;\; g_{tt} = - c_3^2 ( {\cal R}_1 - r_1)^{2 q_2} \cos{\alpha _2} < 0 \; .
$$
Since $k<0$, the inequality (\ref{ter2}) is equivalent to
$$
\left[ (1+\sin{\alpha _1})\cos{\alpha _2} + (1-\sin{\alpha _2})\cos{\alpha _1}  \right] \leq 0 \; ,
$$
but this condition cannot be satisfied if $\cos{\alpha _1} >0$ and $\cos{\alpha _2} > 0$.

\section{Shells}
In this section we construct and analyze the spacetimes corresponding to one or more concentric shells matched to an exotic exterior, and 
explicitly show that the DEC is violated. 
As we shall see, the interior metric can be diagonalized, so the corresponding hypothesis of the previous section trivially holds. However, the distribution of matter is not regular and, in particular, the function $q$ is not continuous at $r={\cal R}_1$, as we have assumed in the previous section. Nevertheless, one can take into account the fact that we
are dealing with shells, make minor changes on the proof above given, and see that the final result still holds. 

For the discussions on energy conditions it is convenient to introduce a normalized base:
$$
\hat{r}^a = ( \partial _r )^a \;\;\; , \;\;\; \hat{\theta }^a = g_{\theta \theta}^{-1/2}( \partial _{\theta} )^a \;\;\; , \;\;\; 
\hat{z}^a = g_{zz}^{-1/2} ( \partial_ z )^a \;\;\; , \;\;\; \hat{t}^a = |g_{tt}|^{-1/2}( \partial _ t )^a \; .
$$
As in the previous Section, we choose $k<0$.
\subsection{One shell}
In this subsection we analyze the case in which there is just one shell of radius, say, ${\cal R}_0$. The interior metric must be flat, otherwise it would be singular. It can be seen that, if the exterior metric is exotic, the shell must violate the DEC. We leave the proof of this statement for the following subsection. Thus, the exterior geometry of a physically reasonable shell is described by a Levi-Civita metric. The regularity at the axis plus the matching conditions at the shell imply that both the exterior and interior metric can be {\em simultaneously} diagonalized, we use this fact to write
\begin{eqnarray}
ds ^2 & = & dr^2 + r^2 d\theta^2 +  dz^2 -  dt^2 \;\;\; , \;\;\;  0 \leq r \leq {\cal R}_0 \;,  \label{int} \\
ds ^2 & = & dr^2 + R_0^2 \left( \frac{r - r_0}{R_0 - r_0} \right)^{-4(\Delta -1)/(\Delta ^2 +3)} d\theta^2 + \left( \frac{r - r_0}{R_0 - r_0} \right)^{4(\Delta +1)/(\Delta ^2+3)}  dz^2 +  \nonumber \\
& & - \left( \frac{r - r_0}{R_0 - r_0} \right)^{2(\Delta ^2-1)/(\Delta ^2+3)}  dt^2 \;\;\; , \;\;\; {\cal R}_0 \leq r \;.  \label{ext1}
\end{eqnarray}
In the above expressions (and for the rest of the paper) we assume, for simplicity, ${\cal R}_0>r_0$; the analysis for ${\cal R}_0<r_0$ is similar, and one can see that the results that we are interested on also hold in that case. We now write down the components of $T_{ab}$ that are obtained when the metric defined by Eqs. (\ref{int},\ref{ext1}) is used to evaluate the l.h.s.
of Einstein equation, $G_{ab} = 8\pi T_{ab}$. We shall need them in the next subsection. The results are
\begin{eqnarray}
\label{tab1}
T_{ab}\hat{\theta }^a \hat{\theta }^b & = &   \frac{(\Delta +1)^2}{8 \pi ({\cal R}_0 - r_0)(\Delta ^2+3)} \delta(r - {\cal R}_0) \; \\
\label{tab2}
T_{ab} \hat{z}^a \hat{z}^b & = & - \frac{2{\cal R}_0 (\Delta +1) - r_0 (\Delta ^2 +3)}  { 8 \pi {\cal R}_0 ({\cal R}_0 - r_0)(\Delta ^2 +3)} \delta(r - {\cal R}_0) \; ,\\
\label{tab3} 
T_{ab} \hat{t}^a\hat{t}^b & = & \frac{{\cal R}_0 (\Delta ^2-1) - r_0 (\Delta ^2 +3)}  {8 \pi {\cal R}_0 ({\cal R}_0 - r_0)(\Delta ^2 +3)} \delta(r - {\cal R}_0)  \;,
\end{eqnarray}
and the trace of this stress - energy tensor is
$$
T = \frac{r_0\delta{(r - {\cal R}_0)}}{ 4\pi {\cal R}_0({\cal R}_0 - r_0)} \;,   \label{trace1}
$$
so $r_0 = 0 \Rightarrow T=0$. That is, though $r_0$ locally corresponds to a simple shift in the $r$ coordinate, it provides non trivial information about the invariant $T$. Thus, one cannot choose $r_0 =0$ without loss of generality. Something similar led to the belief that a rotating cylinder can only exist in general relativity for ``incoherent (traceless) matter'' \cite{fre} (more on rotating cylindrical shells can be found in \cite{cle}).

The SEC, $T_{ab}\hat{m}^a \hat{m}^b + \frac{1}{2}T \geq 0$ for all unit timelike $\hat{m}^a$, implies in this case (choosing $\hat{m}^a = \hat{t}^a$)
$$
T_{ab}\hat{t}^a \hat{t}^b + \frac{1}{2}T =\frac{(\Delta ^2-1)}{8\pi (\Delta ^2+3)({\cal R}_0-r_0)} \delta (r - {\cal R}_0) \geq 0\;,
$$
and it is satisfied if and only if $\Delta ^2 > 1 $ (the inequality must be reversed if ${\cal R}_0 < r_0$). It is easy to see that this is a general property of Levi-Civita metrics, e.g., it does not depend on the fact that we are dealing with shells \cite{gle} (see the last Section for further comments on the SEC).

Similarly, as a necessary condition for the WEC, we have
\begin{equation}
T_{ab}\hat{t}^a \hat{t}^b + T_{ab}\hat{z}^a \hat{z}^b = \frac{(\Delta +1)(\Delta -3)}{8\pi (\Delta ^2+3)({\cal R}_0-r_0)} \delta(r-{\cal R}_0) \geq 0   \; .\label{weca}
\end{equation}
In Sec. IV we saw that the WEC $\Rightarrow |g_{tt}|\geq g_{zz}$ under general conditions. Let us explicitly check this property: writing
$$
g_{tt} + g_{zz} = \left( \frac{r-r_0}{{\cal R}_0-r_0} \right) ^{2(\Delta ^2-1)/(\Delta ^2+3)} \left[ 1- \left( \frac{r-r_0}{{\cal R}_0-r_0} \right)^{-2(\Delta +1)(\Delta -3)/(\Delta ^2+3)} \right] \; ,
$$
we notice that Eq. (\ref{weca}) $\Rightarrow g_{tt} + g_{zz} \geq 0$.

\subsection{Concentric shells}

We now consider a model consisting of two concentric shells. As discussed in the previous subsection, assuming that the interior of the innermost shell is empty and regular, and that it is made out of matter and (or) fields satisfying the DEC, the metric between the innermost and outermost shells must be taken to be of the Levi-Civita form. Our problem then is to analyze the matching of the Levi-Civita metric (\ref{ext1}) with a metric of the form given by (\ref{exotic1},\ref{exotic2},\ref{exotic3},\ref{exotic4}), through a singular shell at $r= {\cal R}_1 > {\cal R}_0$. We obtain
\begin{eqnarray*}
g_{\theta \theta} & = & {\cal R}_0^2 \left( \frac{{\cal R}_1 - r_0}{{\cal R}_0-r_0} \right)^{-4(\Delta -1)/(\Delta ^2 +3)}  \left( \frac{r - r_1}{{\cal R}_1-r_1}  \right)^{2 q_1} \; , \\ 
g_{zz} & = & \frac{1}{\cos{\alpha }} \left( \frac{{\cal R}_1 - r_0}{{\cal R}_0 - r_0} \right)^{4(\Delta +1)/(\Delta ^2+3)} \left( \frac{ r - r_1}{{\cal R}_1-r_1}  \right)^{2 q_2}   \cos{\left[ 2k \ln (r - r_1) + 2 \phi_1) \right]} \; ,  \\
g_{tt} & = & - \frac{1}{\cos{\alpha }}  \left( \frac{{\cal R}_1 - r_0}{{\cal R}_0-r_0} \right)^{2(\Delta ^2-1)/(\Delta ^2+3)} \left( \frac{r - r_1}{{\cal R}_1-r_1}  \right)^{2 q_2} \cos{\left[ 2k \ln (r - r_1) - 4k \ln ( {\cal R}_1 - r_1) - 2 \phi_1 ) \right]} \; ,  \\
g_{tz} & = & \frac{1}{\cos{\alpha }}  \left( \frac{{\cal R}_1 - r_0}{{\cal R}_0-r_0} \right)^{(\Delta +1)^2/(\Delta ^2+3)}  \left( \frac{r - r_1}{{\cal R}_1-r_1}  \right)^{2 q_2} \sin{\left[ 2k \ln (r - r_1) - 2k \ln ({\cal R}_1 - r_1) \right] } \; ,  
\end{eqnarray*}
with $\alpha = 2k\ln{({\cal R}_1 - r_1)} + 2\phi _1$ such that $\cos{\alpha} > 0$. 

The components of the energy-momentum tensor for the innermost shell are given by (\ref{tab1},\ref{tab2},\ref{tab3}), and for the outermost one by
\begin{eqnarray*}
T_{ab} \hat{\theta } ^a \hat{\theta }^b & = & \frac{1}{8\pi } \left[  \frac{2(1-s\sqrt{1+3k^2})}{3({\cal R}_1 - r_1)}  -    \frac{(\Delta +1)^2}{(\Delta ^2+3)({\cal R}_1 - r_0)}  \right] \delta(r - {\cal R}_1) \; , \\
T_{ab} \hat{z}^a \hat{z}^b & = & \frac{1}{8\pi } \left[ \frac{1}{3({\cal R}_1 - r_1)} \left( 2+s\sqrt{1+3k^2} + 3k\tan{\alpha }   \right) - 
 \frac{(\Delta -1)^2}{(\Delta ^2+3)({\cal R}_1 - r_0)}  \right] \delta(r - {\cal R}_1) \; , \\
T_{ab} \hat{t}^a \hat{t}^b & = & \frac{1}{8\pi } \left[  \frac{1}{3({\cal R}_1 - r_1)} \left( -2-s\sqrt{+3k^2} + 3k\tan{\alpha }   \right) +  
\frac{4}{(\Delta ^2+3)({\cal R}_1 - r_0)}  \right] \delta( r - {\cal R}_1) \; , \\
T_{ab} \hat{t}^a \hat{z}^b & = &   \frac{ - k }{8\pi ({\cal R}_1 - r_1) \cos{\alpha }} \delta(r - {\cal R}_1)  \; , 
\end{eqnarray*}
whereas its trace is
$$
T= \frac{ (r_1 - r_0) }{4\pi ({\cal R}_1-r_1) ({\cal R}_1-r_0)} \delta(r - {\cal R}_1)  \; , 
$$
and, similar to what happens for a single shell, $r_1=r_0 \Rightarrow T=0$.

As a necessary condition for the WEC on the outermost shell,
\begin{equation}
T_{ab}\hat{t}^a\hat{t}^b + T_{ab}\hat{z}^a\hat{z}^b = \frac{1}{8\pi} \left[ \frac{2k\tan{\alpha}}{({\cal R}_1-r_1)} - \frac{(\Delta +1)(\Delta -3)}{(\Delta ^2+3)({\cal R}_1-r_0)} \right] \delta (r-{\cal R}_1) \geq 0  \label{wecb} \; .
\end{equation}

Combining Eq. (\ref{wecb}) with the analogous equation for the innermost shell, Eq. (\ref{weca}), we obtain
\begin{equation}
{\cal I}_1 = \int _0^{{\cal R}_1} g^{1/2}  (T_{ab}\hat{t}^a\hat{t}^b + T_{ab}\hat{z}^a\hat{z}^b) = \frac{{\cal R}_0 ({\cal R}_1 - r_0) k\tan {\alpha}}{4\pi ({\cal R}_0-r_0)({\cal R}_1 - r_0)} \geq 0  \label{int1}  \; .
\end{equation}
The other integral we need is
\begin{equation}
{\cal I}_2 = \int _0^{{\cal R}_1} g^{1/2}  T_{ab}\hat{t}^a\hat{z}^b = \frac{ - k {\cal R}_0 ({\cal R}_1-r_0 )}{8\pi ({\cal R}_0-r_0)({\cal R}_1-r_0)  \cos{\alpha }} \; , 
  \label{int2}
\end{equation}
which is also positive, not because of any energy condition, but rather due to the range of the constants that appear on it. Finally, 
the DEC implies that ${\cal I}_1 \geq 2{\cal I}_2$, an inequality that cannot be satisfied if $k <0$, and both (\ref{int1}) and (\ref{int2}) are positive. Therefore, as stated, the DEC is necessarily violated if the matching to the exotic metric is non trivial.

\section{Final comments}

The possibility of the existence of an exotic metric associated with a
physical source is of course very intriguing. In a general sense, it
would amount to some form of ``frame dragging'', that results from the
presence of a momentum flux in the source, such that the stress -
energy - momentum tensor cannot be diagonalized in general, in some way
reminiscent of the frame dragging effect for source endowed with
rotation. Unfortunately, in all the examples analyzed in this paper, we have not been able to construct such a source, if we also impose the dominant energy condition. We have also constructed a general proof of nonexistence of sources satisfying the usual physical requirements, but only under some restrictions. In particular, the  main assumption of the proof given in Section IV is that there is a killing vector field that is {\em everywhere} spatial in the source region, and another one that is {\em everywhere} timelike, in the same region. It applies, in particular, to sources that have small relative (non trivial) flux of momentum, but it is not the general case. For example, the exotic metrics do not satisfy this condition (though, of course, they can not be used as sources because they are not regular at the axis). It would be interesting to have some result concerning sources that do not satisfy this hypotheses, either showing that they can generate exotic metrics, or extending the present proof to those
 cases.

The other important assumption in Sec. IV is that the matter satisfies the DEC. This is usually considered to be a physically reasonable assumption, satisfied by non tachyonic matter, in particular, by topological defects. We notice, however, that some of the latter violate another energy condition, the so called strong energy condition (SEC). There are different ways of realizing that sources for a given spacetime must violate it. For example (following the notation of (\ref{met2}) with $D=0$), if one looks at the geodesic equation, one notices that $A' <0$ corresponds to a ``repulsive'' gravitational field, a situation in which one would suspect that the SEC is being violated. This is indeed the case, the components of the Ricci tensor are
\begin{eqnarray*}
R_{ab}\hat{t}^a \hat{t}^b & = & \frac{1}{2} \left[ A' e^{(A+B+C)/2} \right]' e^{-(A+B+C)/2} \; , \\
R_{ab} \hat{z}^a \hat{z}^b& = & - \frac{1}{2} \left[ B' e^{(A+B+C)/2} \right]' e^{-(A+B+C)/2} \; , \\
R_{\theta \theta} \hat{\theta }^a \hat{\theta }^b& = & - \frac{1}{2} \left[ C' e^{(A+B+C)/2} \right]' e^{-(A+B+C)/2} \; , \\
R_{ab} \hat{r}^a \hat{r}^b & = & -\frac{1}{2} (A+B+C)^{''} - \frac{1}{4} (A^{'2} + B^{'2} + C^{'2})  \; , 
\end{eqnarray*}
and, thus, a {\em regular} solution of Einstein equations satisfies
\begin{equation}
A' =  16 \pi e^{-(A+B+C)/2} \int_0^r e^{(A+B+C)/2} {\cal SEC}  \label{sec} \; ,
\end{equation}
where ${\cal SEC} = T_{ab}\hat{t}^a \hat{t}^b + T/2$. From Eq. (\ref{sec}) we can see that if $A' <0$,  then ${\cal SEC}<0$ (at a set of finite measure), and the SEC is violated.

In general, topological defects that violate the SEC have a global symmetry. For example, following the notation of Section III, for a $U(1)$ global string we have 
$$
{\cal SEC} =  - \frac{1}{8} \lambda (R^2 - \eta ^2)^2 \; , 
$$
which is manifestly negative, so the SEC is {\em everywhere} violated. Similarly, global monopoles \cite{har} and global vacuumless defects \cite{vil} have repulsive gravitational fields that suggest that the SEC is violated. In the case of gauge defects, it is usually supposed that they satisfy the SEC. But, in such cases, we face the difficulty that ${\cal SEC}$ does not have definite sign, and,
 therefore, we do not know whether at a given point it is positive or not without knowledge of the solution of the field equations. Nevertheless, there might be an interesting exception to the general belief that gauge defects satisfy the SEC, an exception that seems not to have been noticed up to present. Amsterdamski and Laguna have numerically solved the equations describing a gravitating superconducting string \cite{lag} and, interestingly, if one observes figure (7) of that paper, one notes that $g_{tt}$ clearly has a local minimum. According to the Eq. (\ref{sec}),  this means that superconducting strings violate the SEC.

\section*{Acknowledgments}

This work was supported in part by funds of the University of C\'ordoba, and grants from CONICET and CONICOR (Argentina). R.J.G. is a member of CONICET. M.H.T. acknowledges financial support from CONICOR and CONICET.

\end{document}